\documentclass[conference]{IEEEtran}
\IEEEoverridecommandlockouts
\usepackage{cite}
\usepackage{citesort}
\usepackage{graphicx}
\usepackage{amsmath}
\usepackage{amsfonts}
\usepackage{xcolor}
\usepackage{amssymb,bm,upgreek}
\usepackage{algorithm}
\usepackage{algorithmic}
\usepackage{multirow}
\usepackage{transparent}
\usepackage{verbatim}
\usepackage{subfigure}

\usepackage{import}

\DeclareMathOperator\erf{erf}
\DeclareMathOperator*{\E}{\mathbb{E}}

\newcommand{\prob}{\mathbb{P}}
\newcommand{\agg}{\textnormal{agg}}
\newcommand{\s}{\textnormal{s}}
\newcommand{\mole}{\textnormal{molecule}}

\newcommand{\m}{\textnormal{m}}
\newcommand{\self}{\textnormal{self}}

\newcommand{\QS}{\textnormal{QS}}
\newcommand{\D}{\textnormal{D}}
\newcommand{\ab}{\textnormal{ab}}

\newtheorem{remark}{Remark}
\newtheorem{theorem}{Theorem}

\newtheorem{lemma}{Lemma}

\def\BibTeX{{\rm B\kern-.05em{\sc i\kern-.025em b}\kern-.08em
    T\kern-.1667em\lower.7ex\hbox{E}\kern-.125emX}}

\newcommand{\Fr}[1]{{\color{black}{#1}}}
\DeclareUnicodeCharacter{00A0}{ }

\begin{document}

\title{Molecular Communication for Quorum Sensing Inspired Cooperative Drug Delivery}

\author{\IEEEauthorblockN{Yuting Fang, {\it  Member, IEEE}, Saman Atapattu, {\it Senior Member, IEEE},\\
Hazer Inaltekin, {\it  Member, IEEE}, and Jamie Evans, {\it Senior Member, IEEE}\\}
\vspace{-5mm}
\thanks{Y. Fang, S. Atapattu, and J.~Evans are with the Department of Electrical and Electronic Engineering, University of Melbourne, Parkville, VIC 3010, Australia.
Email:\{yuting.fang, saman.atapattu, jse\}@unimelb.edu.au.}
\thanks{H.~Inaltekin is with the School of Engineering, Macquarie University, North Ryde, NSW 2109, Australia.
Email: hazer.inaltekin@mq.edu.au.}
\thanks{The work of Y. Fang is supported by the Doreen Thomas Fellowship at the University of Melbourne.}
}

\author{\IEEEauthorblockN{Yuting Fang, {\it  Member, IEEE}, Stuart T. Johnston,
Matt Faria, Xinyu Huang, {\it  Student Member, IEEE},\\ Andrew W. Eckford, {\it Senior Member, IEEE}, and Jamie Evans, {\it Senior Member, IEEE}\\}

\thanks{Y. Fang and J.~Evans are with the Department of Electrical and Electronic Engineering, S.T. Johnston is with the School of Mathematics and Statistics and
Systems Biology Laboratory, Department of Biomedical Engineering, and M. Faria is
with the Department of Biomedical Engineering and ARC Centre of Excellence in Covergent Bio-Nano Science and Technology, The University of Melbourne, Parkville, VIC 3010, Australia. (e-mail:yuting.fang@unimelb.edu.au;stuart.johnston@unimelb.edu.au; matthew.faria@unimelb.edu.au;jse@unimelb.edu.au).}
\thanks{X. Huang is with the School of Engineering, Australian
National University, Canberra, ACT 2600, Australia (e-mail:
xinyu.huang1@anu.edu.au).}
\thanks{Andrew W. Eckford is with the Department of Electrical Engineering and
Computer Science, York University, Toronto, ON M3J 1P3, Canada (e-mail:
aeckford@yorku.ca).}
\thanks{Yuting Fang’s work was supported by the Doreen
Thomas Postdoctoral Fellowship at the University of Melbourne, Stuart T.
Johnston’s work was supported by the Australian Research Council
(project no. DE200100998), Matthew Faria’s work was supported
by a gift from the estate of R\'{e}jane Louise Langlois.}
}

\maketitle
\begin{abstract}
\Fr{A cooperative drug delivery system is proposed, where quorum sensing (QS), a density-dependent bacterial behavior coordination mechanism, is employed by synthetic bacterium-based nanomachines (B-NMs) for controllable drug delivery. In our proposed system, drug delivery is only triggered when there are enough QS molecules, which in turn only happens when there are enough B-NMs. This makes the proposed system can be used to achieve a high release rate of drug molecules from a high number of B-NMs when the population density of B-NMs may not be known.} Analytical expressions for i) the expected activation probability of the B-NM due to randomly-distributed B-NMs and ii) the expected aggregate absorption rate of drug molecules due to randomly-distributed QS activated B-NMs are derived. Analytical results are verified by particle-based simulations. The derived results can help to predict and control the impact of environmental factors (e.g. diffusion coefficient and degradation rate) on the absorption rate of drug molecules since rigorous diffusion-based molecular channels are considered. {Our results show that the activation probability at the B-NM increases as this B-NM is located closer to the center of the B-NM population and the aggregate absorption rate of the drug molecules non-linearly increases as the population density increases.}
\end{abstract}

\IEEEpeerreviewmaketitle

\section{Introduction}\label{sec:intro}

Recently, researchers have expressed interest in designing drug delivery systems based on the molecular communication (MC) paradigm, in which drug carriers are modeled as transmitters (TXs), diseased cells are modeled as receivers (RXs), and drugs' motion is modeled as a random channel; see \cite{9027862}. On the other hand, a survey paper \cite{8735961} has proposed a cooperative drug delivery scenario where a cluster of bacterium-based nanomachines (B-NMs) releases drugs to a targeted diseased cell site. 
B-NMs are nanomachines that are realized through the reuse and reengineering of bacteria, and are envisioned to have bacterial capabilities such as sensing, processing, communication, and group coordination \cite{7060516}. 
However, \cite{8735961} has not theoretically analyzed system performance and has not considered group behavior coordination within B-NMs.

Experimental and theoretical studies have shown that the total drug dosage as well as the drug absorption rate by the diseased cells are critical factors in the healing process \cite{LEE2015158}. 
{One way to achieve a high drug release rate is for B-NMs to ``synchronize'' their release of drug molecules by releasing them only when the local population density of B-NMs is high. However, in general, the population density of local B-NMs is uncertain to an individual B-NM. For example, when B-NMs are engineered from bacteria, they likely undergo cell division and cell death over the lifetime of the application \cite{8735961}. Also, B-NMs have different arrival times to the target site once they are injected. Hence, an efficient method to coordinate density-dependent group behavior is needed to estimate the population density of local B-NMs and subsequently trigger drug delivery at a high population density.}

In nature, quorum sensing (QS) is a common method for many species of bacteria to estimate the local density of other bacteria and trigger certain behaviors when the density is high. In QS, bacteria coordinate their group behavior based on the emission and reception of signalling QS molecules.
Inspired by \cite{8735961} and QS in nature, we propose a QS-like cooperative drug delivery system. 
In this system, QS is used by synthetic B-NMs to estimate the local population density and subsequently {deliver drugs when the population density of B-NMs is sufficiently high}. 
The general process of the proposed system is as follows. The B-NM that reaches a target site (e.g. a tumor) will release QS molecules to the environment to show its presence to other B-NMs. The B-NMs will assess the population density of B-NMs in their local environment by detecting the concentration of QS molecules in the environment. When the number of signalling molecules received at a B-NM is over a threshold (i.e., the population density is above a minimal level for drug delivery), the B-NM is activated for cooperative response and begins delivering drug molecules to the environment. 
In doing so, a cluster of B-NMs will deliver drugs when the population density is high, which maximizes the efficacy of drug treatment and also importantly reduces side effects. 

We make the following novel contributions: {We for the first time propose a QS inspired cooperative drug delivery system and provide performance analysis by considering diffusive MC channels among different B-NMs and the ones between B-NMs and the delivery target site}\footnote{{Several papers such as \cite{Abadal2011,9145736} also have considered QS as an efficient bio-inspired mechanism for achieving global synchronisation among a cluster of nanomachines in a nanonetwork, but they did not provide performance analysis by considering QS molecules' emission, propagation, and reception.}}.
In the proposed system, a cluster of B-NMs is randomly distributed in a three-dimensional (3D) environment. Since bacteria emit molecules sporadically in reality, we assume that each B-NM continuously emits molecules at random times.
We for the first time develop an analytical model to predict i) the expected absorption rate of drug molecules at a RX due to continuous emission of molecules at one TX, ii) the expected activation probability of the B-NM due to the emission of QS molecules from a cluster of randomly-distributed B-NMs, and iii) the expected aggregate absorption rate of drug molecules at a RX due to continuous emission of drug molecules at a cluster of randomly-distributed activated B-NMs.  
This model allows us to interrogate the behavior and likely performance of drug delivery systems using QS coordinated B-NMs to treat diseased cells.

\vspace{-3mm}
\section{System Model}\label{sec:system model}
\begin{figure}[!t]
\centering
\includegraphics[width=0.75\columnwidth]{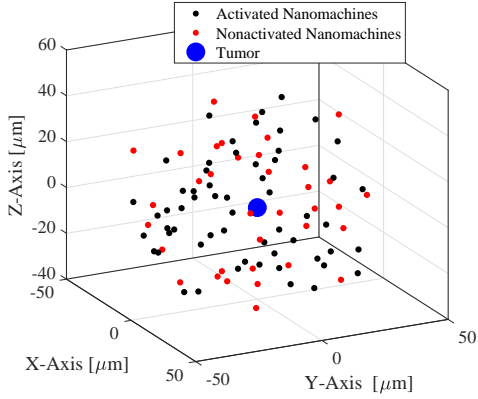}
\vspace{-3mm}
\caption{An example illustration of the system topology: A simulation realization of a 3D PPP of randomly-distributed B-NMs and the resulting activated B-NMs for drug delivery with QS coordination. $R_1=50\,{\mu}\m$, $\lambda=2\times10^{-4}$ B-NMs per ${{\mu}\textrm{m}^{3}}$; see other simulation details in Sec. \ref{sec:Numerical}. The black and red dots represent activated and nonactivated B-NMs, respectively.}
\label{fig:model}
\vspace{-5mm}
\end{figure}

We consider an unbounded 3D environment in which a population of B-NMs use QS to coordinate their group activity for drug delivery to a target site (e.g. a tumor)\footnote{{In a drug delivery system, the B-NMs are either injected directly in the vicinity of the target site of a patient which minimizes the side effects to healthy cells, or intravenously propagate through the bloodstream toward the target location, e.g., extravasated from the blood into the interstitial tissue near the diseased cells \cite{LEE2015158}. In this work, we assume the B-NMs have arrived near the tumor, e.g., by direct injection or via the capillaries.}}. An example of the system topology is shown in Fig.~\ref{fig:model}. B-NMs are randomly located up to a radius $R_1$ from the environment center $(0,0,0)$ at a density\footnote{\Fr{
In reality, colony fitness may detrimentally affect the population size of B-NMs, e.g., the metabolic costs for B-NMs might force B-NMs into quiescence. For tractable analysis, we do not consider the varying population size of B-NMs due to the colony fitness. This assumption is reasonable when B-NMs have the capacity to acquire energy, e.g., a motor protein can acquire energy from ATP hydrolysis [2], or B-NMs store enough energy to perform their functionalities. Moreover, B-NMs may be not able to reproduce to result in colony growth because they are artificially engineered.}} $\lambda$. 
The location of B-NMs are distributed according to a 3D Poisson point process (PPP)\footnote{{We assume that the B-NMs do not move over time after being randomly placed in each realization of the 3D PPP. This assumption is reasonable for three reasons: i) B-NMs are nanomachines that are realized through the reuse and reengineering of bacteria. There are some non-motile bacteria, e.g., coliform and streptococci; ii) When B-NMs swim very slowly, the mobile case can be well approximated by the non-mobile case; and iii) In fact, bacteria often keep stationary when cooperating, e.g., when forming a stable biofilm. Thus, it is reasonable to assume that B-NMs keep stationary during the application of cooperative drug delivery.}}; {see Appendix \ref{proof:realization} for more details about PPP}. 
We denote $\vec{x}_i$ as the location of the center of the $i$th B-NM. We denote $\Phi\left(\lambda\right)$ as the random set of B-NM locations. We model the $i$th B-NM as a point TX that releases QS molecules to the environment and also as a spherical passive RX to detect QS molecules from the environment. We denote the radius of each B-NM by $R_0$. 
B-NMs perfectly count molecules if they are within the volume of B-NM. The tumor is located at $\vec{x}_\ab$, and we model it as a spherical absorbing RX with radius $R_\ab$. 
The tumor perfectly absorbs drug molecules that hit the surface of the tumor.
We present a basic conceptual schematic design of the synthetic B-NM that can execute the drug delivery task based on \cite{8735961} in Fig. \ref{fig:model1}. We next detail the behaviors of the B-NMs, tumor, QS molecules, and drug molecules.


\emph{1) QS molecules' emission, propagation, and reception among B-NMs}:
The $i$th B-NM continuously emits {molecules of the type $A_{\QS}$} from $\vec{x}_i$ at random times from time $t_0=0$ {according to an independent and identically distributed random process} with constant {mean rate} $q_{\QS}$ ${\mole}/{\s}$.
All $A_{\QS}$ molecules diffuse independently with a constant diffusion coefficient $D_{\QS}$. Over time, they can degrade into a form that cannot be detected by the B-NMs, i.e., $A_{\QS}\overset{k_{\QS}}\rightarrow\emptyset$, where $k_{\QS}$ is the reaction rate constant in $\s^{-1}$. We write the number of molecules observed at the $i$th B-NM at time $t_1$ as $N_{\agg}^{\dag}\left(\vec{x}_i,t_1|\lambda,t_0\right)=\sum_{\vec{x}_j\in\Phi\left(\lambda\right)}N\left(\vec{x}_i,t_1|\vec{x}_j,t_0\right)$, where $N\left(\vec{x}_i,t_1|\vec{x}_j,t_0\right)$ is the number of molecules observed at the $i$th B-NM at time $t_1$ due to the continuous emission at the $j$th B-NM since time $t_0=0$. The means of $N_{\agg}^{\dag}\left(\vec{x}_i,t_1|\lambda,t_0\right)$ and $N\left(\vec{x}_i,t_1|\vec{x}_j,t_0\right)$ are denoted by $\overline{N}_{\agg}^{\dag}\left(\vec{x}_i,t_1|\lambda,t_0\right)$ and $\overline{N}\left(\vec{x}_i,t_1|\vec{x}_j,t_0\right)$, respectively.

\emph{2) Activation and drug molecules' emission at B-NMs}:
Inspired by the threshold-based strategy of QS, we assume that the $i$th B-NM is activated for the delivery of drug molecules if $N_{\agg}^{\dag}\left(\vec{x}_i,t_1|\lambda,t_0\right)>\eta$, where $\eta$ is a decision threshold and {$N_{\agg}^{\dag}\left(\vec{x}_i,t_1|\lambda,t_0\right)$ is the instantaneous observation at the $i$th B-NM at the time instant $t_1$}. 
If the $i$th B-NM is activated, the B-NM releases drug molecules $A_{\D}$ to the environment from $\vec{x}_i$ at random times since time $t_1$ according to an independent random process with {constant\footnote{{The assumption of a constant release rate for drugs is reasonable due to that i) B-NMs may be able to produce drug molecules by themselves after being programmed via sophisticated synthetic bioengineering \cite{Duong2019} and ii) B-NMs actually do not need to releases drug molecules with constant mean rate for an infinite time period since we focus on the average absorbing rate of drug molecules in the steady state and it can arrived very quickly, e.g. after 1 s since initial emission of drug molecules, as shown in Fig. 4 and Fig. 5(b).}} mean rate} $q_{\D}$ ${\mole}/{\s}$.

\emph{3) Drug molecules' propagation and reception at tumor}:
All $A_{\D}$ molecules diffuse independently with a constant diffusion coefficient $D_{\D}$. They can degrade into a form that cannot be detected by the tumor, i.e., $A_{\D}\overset{k_{\D}}\rightarrow\emptyset$, where $k_{\D}$ is the reaction rate constant in $\s^{-1}$. If $k_{\D}=0$, {there is no degradation}. We use the aggregate hitting rate (i.e., absorption rate) of the drug molecules at the tumor due to a population of QS coordinated B-NMs at a given time $t$ as the \emph{performance metric} of the QS inspired cooperative drug delivery system\footnote{{Based on \cite{9014054}, the ligand residence time may also affect the performance of MC systems. The analytical study of hitting rate considering the ligand residence time is an interesting topic for future work. For tractable analysis, as in \cite{9027862}, we use the hitting rate, i.e., the absorption rate of a given drug molecule, as the performance metric of the drug delivery system.}}. We write the aggregate hitting rate at the tumor at time $t_2$ as $h_{\agg}\left(\vec{x}_\ab,t_2|t_1\right)=\sum_{\vec{x}_i\in \widetilde{\Phi}}h\left(\vec{x}_\ab,t_2|\vec{x}_i,t_1\right)$, where $\widetilde{\Phi}$ is the subset of the random set $\Phi$ and $\widetilde{\Phi}$ includes all activated B-NMs for drug delivery in $\Phi$. $h\left(\vec{x}_\ab,t_2|\vec{x}_i,t_1\right)$ is the hitting rate of the drug molecules at the tumor at time $t_2$ due to the continuous emission of B-NM located at $\vec{x}_i$ since time $t_1$. The means of $h_{\agg}\left(\vec{x}_\ab,t_2|t_1\right)$ and $h\left(\vec{x}_\ab,t_2|\vec{x}_i,t_1\right)$ are denoted by $\overline{h}_{\agg}\left(\vec{x}_\ab,t_2|t_1\right)$ and $\overline{h}\left(\vec{x}_\ab,t_2|\vec{x}_i,t_1\right)$, respectively.

As shown in Fig. \ref{fig:time} in Appendix \ref{SupplyResults}, the channel responses, i.e., the average number of QS molecules observed $\overline{N}_{\agg}^{\dag}\left(\vec{x}_i,t_1|\lambda,t_0\right)$ or the average absorbing rate of drug molecules $\overline{h}_{\agg}\left(\vec{x}_\ab,t_2|t_1\right)$, reach the steady (asymptotic) state very quickly, e.g. after $1\,\s$ since initial emission of molecules\footnote{{We denote time $t^{\star}$ as the time after which the absorbed drug molecules $\overline{h}_{\agg}\left(\vec{x}_\ab,t_2|t_1\right)$ reach the steady state, i.e.,
$\overline{h}_{\agg}\!\left(\vec{x}_\ab,t_2|t_1\right)\!|_{t_2-t_1>t^{\star}_i}\!\approx\!\lim_{t_2\rightarrow\infty}h_{\agg}\left(\vec{x}_\ab,t_2|t_1\right)$.
Here, we assume each B-NM can carry and release at least $q_{\D}t^{\star}$ drug molecules to make sure the absorbed drug molecules can reach the steady state, where we recall the B-NM releases drug molecules $A_{\D}$ to the environment with the rate $q_{\D}$ ${\mole}/{\s}$.}}. In this work, for convenience, we consider that the $i$th B-NM uses its steady observation\footnote{\Fr{Even bacteria can reach the steady state quickly, we do not consider the short transient times for bacteria to trigger behavior change since the timescale of gene regulation to coordinate behavior is on the order of minutes/hours. For example, based on
\cite{Danino2010,Trovato2014,Surette7046}, the cooperation of bacteria is observed after the signaling molecules diffuse for at least tens of minutes. In this work, for convenience, we consider that the $i$th B-NM uses its steady observation, $\lim_{t_1\rightarrow\infty}N_{\agg}^{\dag}\left(\vec{x}_i,t_1|\lambda,t_0\right)$, to make a decision if being activated to release drug molecules. This assumption is reasonable since bacteria can reach the steady state quickly, especially relative to the timescale of gene regulation for coordinating behaviors.}},
$\lim_{t_1\rightarrow\infty}N_{\agg}^{\dag}\left(\vec{x}_i,t_1|\lambda,t_0\right)$, to make a decision if being activated to release drug molecules.
We also use the steady hitting rate, $\lim_{t_2\rightarrow\infty}h_{\agg}\left(\vec{x}_\ab,t_2|t_1\right)$, as the system performance metric. For compactness, we remove $t_0$, $t_1$, and $t_2$ in all notations in the remainder of this paper since we focus on observations or performance metrics at the asymptotic stage. 


\vspace{-3mm}
\section{Channel Response Due to Continuous Emission}\label{sec:Prob}

In this section, we aim to investigate the asymptotic channel response, i.e., the expected number of molecules observed at a passive RX or the hitting rate of molecules at an absorbing RX, due to continuous emission of molecules at random times from a point TX. 
The analyses of the channel responses lay the foundations for our derivations of the aggregate hitting rate at the tumor due to a cluster of QS coordinated B-NMs. We next obtain the channel responses due to continuous emission in the following theorems.



\begin{theorem}[Passive RX for $|\vec{x}_i -\vec{x}_j|\neq0$]\label{Theorem:continuous,any}
The asymptotic expected number of QS molecules observed for the passive RX with radius $R_0$ centered at $\vec{x}_i$, due to continuous emission with the {mean rate} $q_{\QS}$ from point $\vec{x}_j$ since time $t=0$, $\overline{N}\left(\vec{x}_i|\vec{x}_j\right)$, using uniform concentration assumption, is given by 
\begin{align}\label{point1}
\overline{N}\left(\vec{x}_i|\vec{x}_j\right)
\approx \frac{\exp\left(-\sqrt{\frac{k_{\QS}}{D_{\QS}}}|\vec{x}_i -\vec{x}_j|\right)q_{\QS} R_0^3}{3D_{\QS}|\vec{x}_i -\vec{x}_j|}.
\end{align}
\end{theorem}
The case when the TX is at the center of RX is included since B-NMs also receive the molecules released from themselves.
\begin{theorem}[Passive RX for $|\vec{x}_i -\vec{x}_j|=0$]\label{Theorem:continuous,0}
The asymptotic expected number of QS molecules observed at the passive RX, due to continuous emission with the {mean rate} $q_{\QS}$ from the center of this RX since time $t=0$, $\overline{N}_{\self}=\overline{N}\left(\vec{x}_i|\vec{x}_j\right)|_{|\vec{x}_i -\vec{x}_j|=0}$, is given by
\begin{align}\label{point-circle,self}
\overline{N}_{\self}
= & \frac{q_{\QS}}{D_{\QS} k_{\QS}^{3/2}}\left(D_{\QS} \sqrt{k_{\QS}}-\sqrt{D_{\QS}}\exp\left(-\sqrt{\frac{k_{\QS}}{D_{\QS}}}R_0\right)\right.\nonumber\\
&\times\left.\left(\sqrt{D_{\QS}k_{\QS}}+k_{\QS}R_0\right)\right).
\end{align}
\end{theorem}

\begin{theorem}[Absorbing RX for $|\vec{x}_\ab-\vec{x}_i|\neq0$]\label{Theorem:continuous,any,ab}
The asymptotic expected hitting rate of drug molecules at the absorbing RX with the radius $R_\ab$ centered at $\vec{x}_\ab$, due to continuous emission with {mean rate} $q_{\D}$ from the point $\vec{x}_i$ since time $t=0$, $\overline{h}\left(\vec{x}_\ab|\vec{x}_i\right)$, is given by
\begin{align}\label{point1,ab}
\overline{h}\left(\vec{x}_\ab|\vec{x}_i\right)\!=\! \exp\left(\!\sqrt{\frac{k_{\D}}{D_{\D}}}(R_\ab-|\vec{x}_\ab-\vec{x}_i|)\!\right)\!\frac{q_{\D} R_\ab}{|\vec{x}_\ab-\vec{x}_i|}.
\end{align}
\end{theorem}
\begin{IEEEproof}
The proofs of Theorems \ref{Theorem:continuous,any}, \ref{Theorem:continuous,0}, and \ref{Theorem:continuous,any,ab} are given in Appendix~\ref{Proof of CR}.
\end{IEEEproof}

\vspace{-3mm}
\section{Aggregate Absorption Rate of Drug Molecules with QS Coordination}\label{sec:Prob1}
We use the expected asymptotic aggregate absorption rate of the drug molecules at the tumor due to a population of QS coordinated B-NMs over the spatial random point process $\Phi(\lambda)$, ${\E}_{\Phi}\Big\{\overline{h}_{\agg}\left(\vec{x}_\ab\right)\Big\}$, as the performance metric to quantify the efficacy of the cooperative drug delivery system. We first write the exact expression for ${\E}_{\Phi}\Big\{\overline{h}_{\agg}\left(\vec{x}_\ab\right)\Big\}$ as
\begin{align}\label{agg,abRate,exact}
&\;{\E}_{\Phi}\Big\{\overline{h}_{\agg}\left(\vec{x}_\ab\right)\Big\}
= {\E}_{\Phi}\Big\{\sum_{\vec{x}_i\in \widetilde{\Phi}}\overline{h}\left(\vec{x}_\ab|\vec{x}_i\right)\Big\}\nonumber\\
= &\;{\E}_{\Phi}\Big\{\sum_{\vec{x}_i\in \Phi}\overline{h}\left(\vec{x}_\ab|\vec{x}_i\right)\prob\left(N_{\agg}^{\dag}(\vec{x}_i|\lambda)\geq\eta|\phi(\lambda)\right)\Big\},
\end{align}
where $\prob\left(\cdot\right)$ denotes probability and $\phi(\lambda)$ is a particular realization of $\Phi(\lambda)$. To derive ${\E}_{\Phi}\Big\{\overline{h}_{\agg}\left(\vec{x}_\ab\right)\Big\}$, we first derive
${\E}_{\Phi}\Big\{\prob\left(N_{\agg}^{\dag}(\vec{x}_i|\lambda)\geq\eta|\phi(\lambda)\right)\Big\}$ in the following lemma:
\begin{lemma}\label{lem:prob}
The approximate expression for the expected activation probability at a B-NM centered at a fixed location $\vec{x}_i$ over the spatial point process $\Phi$, ${\E}_{\Phi}\Big\{\prob\left(N_{\agg}^{\dag}(\vec{x}_i|\lambda)\geq\eta|\phi(\lambda)\right)\Big\}$, is
\begin{align}\label{activation, prob}
&\;{\E}_{\Phi}\Big\{\prob\left(N_{\agg}^{\dag}(\vec{x}_i|\lambda)\geq\eta|\phi(\lambda)\right)\Big\}\nonumber\\
= &\; \left(1-\sum_{n=0}^{\eta-1}\frac{1}{n!}\frac{\partial^n \mathcal{L}_{\overline{N}_{\agg}^{\dag}(\vec{x}_i|\lambda)}(-\rho)}{\partial \rho^n}\Bigg|_{\rho=-1}\right),
\end{align}
where
\begin{align}\label{Laplace1}
&\;\mathcal{L}_{\overline{N}_{\agg}^{\dag}(\vec{x}_i|\lambda)}(s)
=\exp\Bigg\{-s\overline{N}_{\self}-\acute{\lambda}\int_{|\vec{r}|=0}^{R_1}\int_{\theta=0}^{2\pi}\int_{\varphi=0}^{\pi}\nonumber\\
&\;\times\left(1-\exp\left(- \frac{s\exp\left(-\sqrt{\frac{k_{\QS}}{
D_{\QS}}}\Omega\left(|\vec{x}_i|,|\vec{r}|,\varphi,\theta\right)\right)q_{\QS} R_0^3}{3D_{\QS}\Omega\left(|\vec{x}_i|,|\vec{r}|,\varphi,\theta\right)}\right)\right)\nonumber\\
&\;\times |\vec{r}|^2 \sin\varphi\,d|\vec{r}|\,d\theta\,d\varphi \Bigg\},
\end{align}
$\Omega\left(|\vec{x}_i|,|\vec{r}|,\varphi,\theta\right)=\sqrt{{|\vec{x}_i|}^2+|\vec{r}|^2+2|\vec{x}_i||\vec{r}|\sin\varphi\cos\theta}$, and $\acute{\lambda}={\left(\lambda\frac{4}{3}\pi R_1^3-1\right)}/{\frac{4}{3}\pi R_1^3}$.
\end{lemma}
\begin{IEEEproof}
See Appendix \ref{proof:Laplace}.
\end{IEEEproof}

\begin{remark}
{To derive \eqref{activation, prob}, we assume ${N}_{\agg}^{\dag}(\vec{x}_i|\lambda)$ is a Poisson RV with mean $\overline{N}_{\agg}^{\dag}(\vec{x}_i|\lambda)$. $N_{\agg}^{\dag}\left(\vec{x_i}|\lambda\right)$ is actually a Poisson binomial RV for the following reasons. We recall that $N_{\agg}^{\dag}\left(\vec{x_i}|\lambda\right)$ is the sum of $N\left(\vec{x_i}|\vec{x_j}\right)$ over $j$. We note that $N\left(\vec{x_i}|\vec{x_j}\right)$ is the sum of the number of molecules observed at the $i$th B-NM at time $t_1$, when these molecules are released from the $j$th B-NM since $t_0=0$. Thus, the observations at the $i$th B-NM due to continuous emission at the $j$th B-NM are not identically distributed since they are released at different times. Therefore, $N\left(\vec{x_i}|\vec{x_j}\right)$ is a Poisson binomial RV since each molecule behaves independently and has a different probability of being observed at $t$ by the $i$th B-NM due to different releasing times. Since $N_{\agg}^{\dag}\left(\vec{x_i}|\lambda\right)$ is the sum of $N\left(\vec{x_i}|\vec{x_j}\right)$, $N_{\agg}^{\dag}\left(\vec{x_i}|\lambda\right)$ is also a Poisson binomial RV. We note that modeling $N_{\agg}^{\dag}\left(\vec{x_i}|\lambda\right)$ as a Poisson binomial RV makes the evaluation of (12) very cumbersome. For tractable performance analysis, we approximate $\overline{N}_{\agg}^{\dag}\!\left(\vec{x_i},\infty|\lambda\right)$ as a Poisson RV. Our simulation results in Sec. V will verify the accuracy of this approximation.}
\end{remark}

Using Lemma \ref{lem:prob}, we next derive an approximated expression for ${\E}_{\Phi}\Big\{\overline{h}_{\agg}\left(\vec{x}_\ab\right)\Big\}$ in the following theorem.
\begin{theorem}\label{the:AggDrug}
The approximate expression for the expected asymptotic aggregate absorption rate of the drug molecules at the tumor due to a population of activated B-NMs with QS coordination, over the spatial random point process $\Phi(\lambda)$, ${\E}_{\Phi}\Big\{\overline{h}_{\agg}\left(\vec{x}_\ab\right)\Big\}$, is
\begin{align}\label{agg,abRate,app}
&\;{\E}_{\Phi}\Big\{\overline{h}_{\agg}\left(\vec{x}_\ab\right)\Big\}\nonumber\\
\approx &\;{\E}_{\Phi}\Big\{\sum_{\vec{x}_i\in \Phi}\overline{h}\left(\vec{x}_\ab|\vec{x}_i\right){\E}_{\Phi}\Big\{\prob\left(N_{\agg}^{\dag}(\vec{x}_i|\lambda)\geq\eta|\phi(\lambda)\right)\Big\}\Big\}\\
\approx &\; q_{\D}\lambda\int_{\alpha=0}^{\pi}\int_{|\vec{x}|=0}^{R_1}\int_{\beta=0}^{2\pi}\mathcal{F}\left(\vec{x},\vec{x}_\ab,\alpha,\beta\right)\nonumber\\
&\;\times\left(1-\sum_{n=0}^{\eta-1}\frac{1}{n!}\frac{\partial^n \mathcal{L}_{\overline{N}_{\agg}^{\dag}(\vec{x}|\lambda)}(-\rho)}{\partial \rho^n}\Bigg|_{\rho=-1}\right)\,d\alpha\, d|\vec{x}| \,d\beta
\end{align}
where
\begin{align}
&\;\mathcal{F}\left(\vec{x},\vec{x}_\ab,\alpha,\beta\right)\nonumber\\
= &\;\exp\left(\sqrt{\frac{k_{\D}}{D_{\D}}}(R_\ab-\Lambda(\vec{x},\vec{x}_\ab))\right)\frac{R_\ab}{\Lambda(\vec{x},\vec{x}_\ab)}
|\vec{x}|^2\sin\alpha
\end{align}
and $\Lambda(\vec{x},\vec{x}_\ab,\alpha,\beta)= \sqrt{|\vec{x}|^2+|\vec{x}_\ab|^2+2|\vec{x}||\vec{x}_\ab|\sin\alpha\cos\beta}$. 
\end{theorem}
\begin{IEEEproof}
The proof of Theorem \ref{the:AggDrug} and the approximation used in (8) are detailed given in Appendix \ref{proof:AggDrug}.
\end{IEEEproof}

\vspace{-3mm}
\section{Numerical Results and Simulations}\label{sec:Numerical}

{In this section, we present simulation and numerical results to assess the accuracy of our derived analytical results and evaluate the performance of our proposed system in the realistic therapeutic scenario. We also reveal the useful impact of environmental parameters on the activation probability and the aggregate absorption rate of drug molecules with QS coordination. The simulation parameters are shown in Table \ref{tab:table1}. The values of those parameters are chosen to be on the same orders as those used in \cite{Danino2010,Trovato2014,Surette7046,Dilanji2012QuorumAA}, so as to better simulate the performance of real systems. In particular, the chosen value of $D_{\QS}$ is the diffusion coefficient of the 3OC6-HSL QS molecules in water at room temperature \cite{Dilanji2012QuorumAA}. The chosen value of $D_{\D}$ is the diffusion coefficient of the ${\textrm N}_{4}$Gd drug molecules \cite{Wang2014}. The volume of B-NM with the chosen radius is approximately equal to the volume of the \emph{V. fischeri} bacteria. We consider an application of our proposed system to a tumor site in the human body. The B-NMs can be injected or extravasated from the cardiovascular system in the tissue surrounding the targeted diseased tumor site \cite{LEE2015158}. Tumor cells may exist in the human body for many years before they start to metastasize \cite{Friberg1997}. When metastasis have often formed, cancer therapies may not be effective. Thus, we assume the size of tumor cells is on the order of ${\mu}\m$ when tumor cells have not started to metastasize and the therapeutic treatment is a practical objective.}

\begin{figure}[!t]
\centering
\includegraphics[height=2.2in]{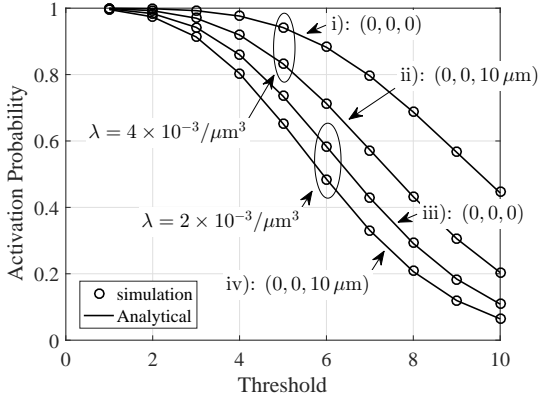}\vspace{-2mm}
\caption{The expected activation probability at the B-NM at the different locations versus threshold due to a population of randomly distributed B-NMs with $R_1=10.60\,{\mu}\m$ for different population densities of B-NMs: Case i) the B-NM is fixed at $(0,0,0)$ with $\lambda=4\times10^{-3}/{{\mu} \textrm{m}^{3}}$, Case ii) the B-NM is fixed at $(0,0,10\,{\mu} \textrm{m})$ with $\lambda=4\times10^{-3}/{{\mu} \textrm{m}^{3}}$, Case iii) the B-NM is fixed at $(0,0,0)$ with $\lambda=2\times10^{-3}/{{\mu} \textrm{m}^{3}}$, and Case iv) the B-NM is fixed at $(0,0,10\,{\mu} \textrm{m})$ with $\lambda=2\times10^{-3}/{{\mu} \textrm{m}^{3}}$.}
\label{SD-opti1}\vspace{-4mm}
\end{figure}

\begin{figure}[!t]
\centering
\includegraphics[height=2.2in]{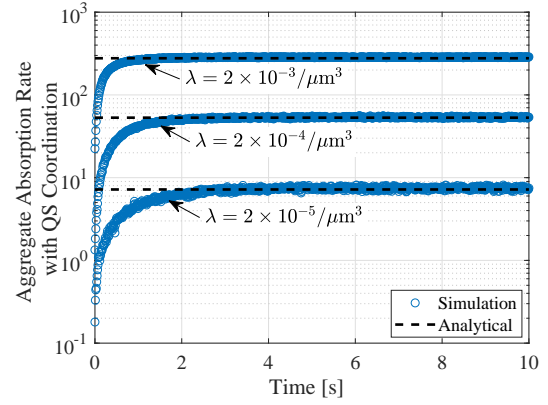}\vspace{-2mm}
\caption{The expected aggregate absorption rate of the drug molecules at the tumor versus time due to a population of QS coordinated randomly distributed B-NMs for different population densities of B-NMs $\lambda=2\times10^{-3}/{{\mu} \textrm{m}^{3}}$, $\lambda = 2\times10^{-4}/{{\mu} \textrm{m}^{3}}$, and $\lambda = 2\times10^{-5}/{{\mu} \textrm{m}^{3}}$. We keep the average number of B-NMs fixed at $10$ and we also consider the tumor located at $(0,0,0)$.}
\label{SD-opti2}\vspace{-4mm}
\end{figure}

We simulate the Brownian motion of QS and drug molecules using a particle-based method \cite{Andrew2004}. In simulations, the locations of bacteria are randomly generated according to a 3D PPP and the times of releasing molecules at each bacterium are generated according to an independent Poisson process. The analytical results in Figs. \ref{SD-opti1} and \ref{SD-opti2} are obtained by \eqref{activation, prob} and \eqref{agg,abRate,app}, respectively. The accuracy of \eqref{point1}, \eqref{point-circle,self}, and \eqref{point1,ab} are validated via Fig. \ref{fig-1} in Appendix \ref{SupplyResults}.

\begin{table}[!t]
\renewcommand{\arraystretch}{1.2}
\centering
\caption{Environmental Parameters}\label{tab:table1}\vspace{-1mm}
\begin{tabular}{c||c|c}
\hline
\bfseries Parameter &  \bfseries Symbol&  \bfseries Value \\
\hline\hline
Radius of B-NM & $R_{0}$ & $0.757\,{\mu}\m$ \\\hline
Radius of tumor & $R_\ab$ & $1\,{\mu}\m$ \\\hline
Diffusion coefficient of $A_{\QS}$ & $D_{\QS}$ & $5.5\times10^{-10}{\m^{2}}/{\s}$\\\hline
Emission rate of $A_{\QS}$ & $q_{\QS}$ & $1 \times10^{4}{\mole}/{\s}$\\\hline
Reaction rate constant of $A_{\QS}$ & $k_{\QS}$ & $1\times10^{1}/{\s}$\\\hline
Diffusion coefficient of $A_{\D}$ & $D_{\D}$ & $8\times10^{-11}{\m^{2}}/{\s}$\\\hline
Emission rate of $A_{\D}$ & $q_{\D}$ & $5 \times10^{2}{\mole}/{\s}$\\\hline
Reaction rate constant of $A_{\D}$ & $k_{\D}$ & $8\times10^{-1}/{\s}$\\\hline
\end{tabular}
\vspace{-3mm}
\end{table}

In Fig. \ref{SD-opti1}, we plot the expected activation probability at the B-NM versus threshold due to a population of randomly distributed B-NMs for different cases, which are specified in the caption. By comparing Case i) with Case iii) (or Case ii) with Case iv)), we see that the activation probability increases as the population density of B-NMs increases. By comparing Case i) with Case ii) (or Case iii) with Case iv)), we see that the activation probability at the B-NM increases as this B-NM is located closer to the center of the B-NM population.

In Fig. \ref{SD-opti2}, we plot the expected aggregate absorption rate of the drug molecules versus time for different population densities of B-NMs. We see that the aggregate absorption rate of the drug molecules increases as the population density increases, but the increase in the aggregate absorption rate is not linearly proportional to the increase in the population density increases. This is expected since \eqref{agg,abRate,app} shows that the aggregate absorption rate is not linearly related to the population density.

\vspace{-3mm}
\section{Conclusions}\label{sec:con}

In this letter, we for the first time modeled a QS inspired cooperative drug delivery system. We also derived analytical expressions for the expected activation probability of the B-NM due to the emission of QS molecules from a cluster of randomly-distributed B-NMs and the expected aggregate absorption rate of drug molecules due to randomly-distributed QS coordinated B-NMs. Our model and results have important relevance for the future design of QS coordinated B-NM drug delivery systems and may be used to predict and optimize their performance in the future. Interesting future research includes incorporating additional relevant biological factors, e.g. the biased motion of B-NMs over time due to chemotaxis and analytically assessing their impacts on drug absorption rate. \Fr{Modeling the time-varying population size of B-NMs due to colony fitness and integrating them into the current analytical framework is also an interesting open problem. To solve this problem, we could update the population size of B-NMs over different time rounds. In each round, we determine B-NMs' death and reproduction probabilities in terms of their nutrient, resource, and energy levels.}

\vspace{-2mm}

\begin{thebibliography}{10}
\providecommand{\url}[1]{#1}
\csname url@samestyle\endcsname
\providecommand{\newblock}{\relax}
\providecommand{\bibinfo}[2]{#2}
\providecommand{\BIBentrySTDinterwordspacing}{\spaceskip=0pt\relax}
\providecommand{\BIBentryALTinterwordstretchfactor}{4}
\providecommand{\BIBentryALTinterwordspacing}{\spaceskip=\fontdimen2\font plus
\BIBentryALTinterwordstretchfactor\fontdimen3\font minus
  \fontdimen4\font\relax}
\providecommand{\BIBforeignlanguage}[2]{{%
\expandafter\ifx\csname l@#1\endcsname\relax
\typeout{** WARNING: IEEEtran.bst: No hyphenation pattern has been}%
\typeout{** loaded for the language `#1'. Using the pattern for}%
\typeout{** the default language instead.}%
\else
\language=\csname l@#1\endcsname
\fi
#2}}
\providecommand{\BIBdecl}{\relax}
\BIBdecl

\bibitem{9027862}
T.~N. Cao, A.~Ahmadzadeh, V.~Jamali, W.~Wicke, P.~L. Yeoh, J.~Evans, and
  R.~Schober, ``Diffusive mobile {MC} with absorbing receivers: Stochastic
  analysis and applications,'' \emph{{IEEE} Trans. Mol. Bio. Multi-Scale
  Commun.}, vol.~5, no.~2, pp. 84--99, 2019.

\bibitem{8735961}
T.~{Nakano}, Y.~{Okaie}, S.~{Kobayashi}, T.~{Hara}, Y.~{Hiraoka}, and
  T.~{Haraguchi}, ``Methods and applications of mobile molecular
  communication,'' \emph{Proceedings of the IEEE}, vol. 107, no.~7, pp.
  1442--1456, 2019.

\bibitem{7060516}
I.~F. {Akyildiz}, M.~{Pierobon}, S.~{Balasubramaniam}, and Y.~{Koucheryavy},
  ``{The Internet of Bio-NanoThings},'' \emph{{IEEE} Commun. Mag.}, vol.~53,
  no.~3, pp. 32--40, 2015.

\bibitem{LEE2015158}
B.~K. Lee, Y.~H. Yun, and K.~Park, ``Smart nanoparticles for drug delivery:
  Boundaries and opportunities,'' \emph{Chemical Engineering Science}, vol.
  125, pp. 158 -- 164, 2015.

\bibitem{Abadal2011}
S.~{Abadal} and I.~F. {Akyildiz}, ``Bio-inspired synchronization for
  nanocommunication networks,'' in \emph{Proc. IEEE GLOBECOM}, Dec. 2011, pp.
  1--5.

\bibitem{9145736}
P.~S.~S. {Tissera} and S.~{Choe}, ``Bio-inspired quorum sensing-based
  nanonetwork synchronization using birth-death growth model,'' \emph{IEEE
  Transactions on Communications}, pp. 1--1, 2020.

\bibitem{Duong2019}
M.~T.-Q. Duong, Y.~Qin, S.-H. You, and J.-J. Min, ``Bacteria-cancer
  interactions: bacteria-based cancer therapy,'' \emph{Exp. Mol. Med.},
  vol.~51, no.~12, pp. 1--15, 2019.

\bibitem{9014054}
U.~A.~K. Chude-Okonkwo, B.~T. Maharaj, A.~V. Vasilakos, and R.~Malekian,
  ``Exploring the impact of ligand residence time on molecular communication
  system performance,'' in \emph{Proc. IEEE GLOBECOM}, 2019, pp. 1--6.

\bibitem{Danino2010}
T.~Danino, O.~Mondrag{\'o}n-Palomino, L.~S. Tsimring, and J.~Hasty, ``A
  synchronized quorum of genetic clocks,'' \emph{Nature}, vol. 463, pp.
  326--330, Jan. 2010.

\bibitem{Trovato2014}
A.~Trovato, F.~Seno, M.~Zanardo, S.~Alberghini, A.~Tondello, and A.~Squartini,
  ``Quorum vs. diffusion sensing: {A} quantitative analysis of the relevance of
  absorbing or reflecting boundaries,'' \emph{FEMS Microbiology Lett.}, vol.
  352, no.~2, pp. 198--203, Jan. 2014.

\bibitem{Surette7046}
M.~G. Surette and B.~L. Bassler, ``Quorum sensing in {Escherichia} coli and
  {Salmonella} typhimurium,'' \emph{Proc. Nat. Academy Sci.}, vol.~95, no.~12,
  pp. 7046--7050, Jun. 1998.

\bibitem{Dilanji2012QuorumAA}
G.~E. Dilanji, J.~B. Langebrake, P.~D. Leenheer, and S.~J. Hagen, ``Quorum
  activation at a distance: spatiotemporal patterns of gene regulation from
  diffusion of an autoinducer signal.'' \emph{J. Am. Chem. Soc.}, vol. 134,
  no.~12, pp. 5618--5626, 2012.

\bibitem{Wang2014}
X.~Wang and Others, ``Diffusion of drug delivery nanoparticles into biogels
  using time-resolved micromri,'' \emph{J. Phys. Chem. Lett.}, vol.~5, no.~21,
  pp. 3825--3830, 2014.

\bibitem{Friberg1997}
S.~Friberg and S.~Mattson, ``On the growth rates of human malignant tumors:
  Implications for medical decision making,'' \emph{J. Surgical Oncol.},
  vol.~65, no.~4, pp. 284--297, 1997.

\bibitem{Andrew2004}
S.~S. Andrews and D.~Bray, ``Stochastic simulation of chemical reactions with
  spatial resolution and single molecule detail,'' \emph{Physical Biology},
  vol.~1, no.~3, pp. 135--151, Aug. 2004.

\bibitem{6949026}
A.~{Noel}, K.~C. {Cheung}, and R.~{Schober}, ``A unifying model for external
  noise sources and {ISI} in diffusive molecular communication,'' \emph{{IEEE}
  J. Select. Areas Commun.}, vol.~32, no.~12, pp. 2330--2343, 2014.

\bibitem{Noel2013}
A.~Noel, K.~C. Cheung, and R.~Schober, ``Using dimensional analysis to assess
  scalability and accuracy in molecular communication,'' in \emph{Proc. IEEE
  ICC}, June 2013, pp. 818--823.

\bibitem{6807659}
H.~B. Yilmaz, A.~C. Heren, T.~Tugcu, and C.-B. Chae, ``Three-dimensional
  channel characteristics for molecular communications with an absorbing
  receiver,'' \emph{{IEEE} Commun. Lett.}, vol.~18, no.~6, pp. 929--932, 2014.

\bibitem{8030318}
Y.~Deng, A.~Noel, W.~Guo, A.~Nallanathan, and M.~Elkashlan, ``Analyzing
  large-scale multiuser molecular communication via 3-{D} stochastic
  geometry,'' \emph{{IEEE} Trans. Mol. Bio. Multi-Scale Commun.}, vol.~3,
  no.~2, pp. 118--133, June 2017.

\bibitem{Haenggi:2012:SGW:2480878}
M.~Haenggi, \emph{Stochastic Geometry for Wireless Networks}, 1st~ed.\hskip 1em
  plus 0.5em minus 0.4em\relax MA, NY: Cambridge University Press, 2012.

\end{thebibliography}

\clearpage

\begin{center}
\textsc{\normalsize{Supplementary Information}}\\
\end{center}

\numberwithin{equation}{section}
\begin{appendices}

\section{Supplementary Figure and Numerical Results}\label{SupplyResults}

\begin{figure}[!h]
\centering
\includegraphics[width=0.4 \textwidth]{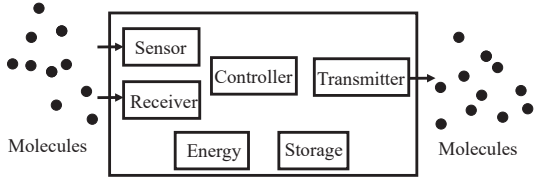}
\vspace{-3mm}
\caption{{A basic conceptual schematic model of the B-NM based on \cite{8735961}. Transmitters are used to emit QS and drug molecules. Sensors are used to detect the presence of QS molecules in the environment. Receivers are used to capture QS molecules. A controller is used to integrate multiple sensory inputs and make decision of activation or not. A storage unit is to store drug and QS molecules. Drug molecules are encapsulated into storage unit. A B-NM may synthesizes QS molecules or captures QS molecules in its environment and stores these molecules. A energy unit is to acquire and expend energy to perform its functionalities.}}
\label{fig:model1}
\vspace{-5mm}
\end{figure}

\begin{figure}[!h]
\centering
\includegraphics[height=2.2in]{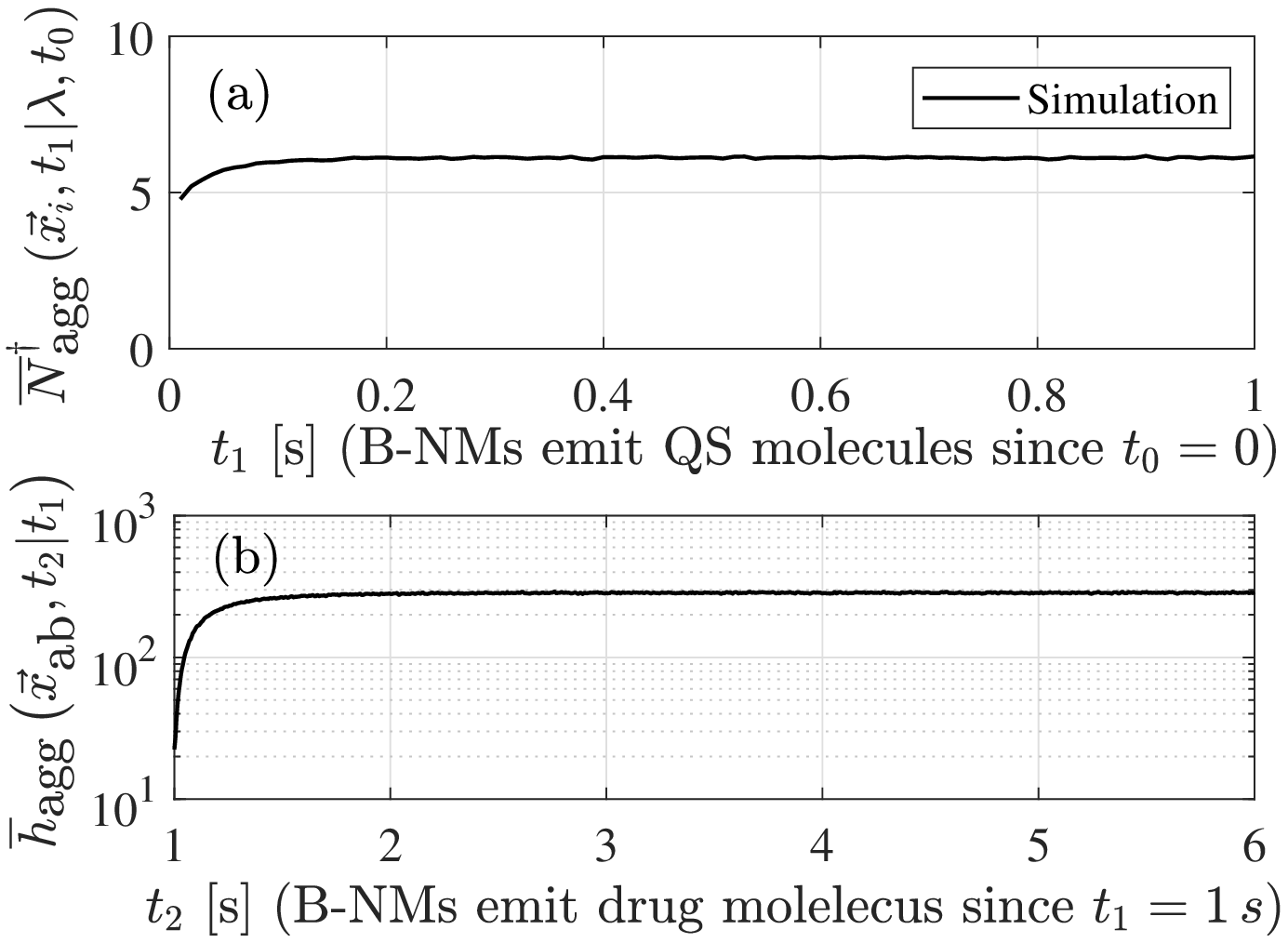}
\vspace{0mm}
\caption{(a): The average number of molecules observed $\overline{N}_{\agg}^{\dag}\left(\vec{x}_i,t_1|\lambda,t_0\right)$ versus time $t_1$ for $\vec{x}_i=(-1.64\,{\mu}\m,-3.59\,{\mu}\m,,9.81\,{\mu}\m)$. (b): The average absorbing rate of molecules $\overline{h}_{\agg}\left(\vec{x}_\ab,t_2|t_1\right)$ versus time $t_2$ for $\vec{x}_\ab = (0,0,0)$. $R_1 = 10.61\,{\mu}\m$, $\lambda=2\times10^{-3}/{{\mu} \textrm{m}^{3}}$; see other simulation details in Sec. \ref{sec:Numerical}.}
\label{fig:time}
\vspace{-2mm}
\end{figure}

\begin{figure}[!h]
\centering
\includegraphics[height=2.2in]{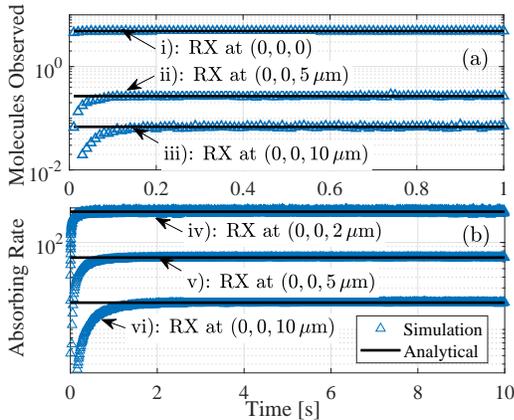}\vspace{-2mm}
\vspace{-2mm}
\caption{The expected number of molecules observed/absorbed by the RX versus time due to the continuous emission of one TX located at $(0,0,0)$.
In Fig. 5(a), we consider passive RXs and three different locations of the RX: Case i) $(0,0,0)$, Case ii) $(0,0,5\,{\mu} \textrm{m})$, and Case iii) $(0,0,10\,{\mu} \textrm{m})$. In Fig. 5(b), we consider absorbing RXs and three different locations of the RX: Case iv) $(0, 0, 2\,{\mu}\m)$, Case v) $(0,0,5\,{\mu}\m)$, and Case vi) $(0,0,10\,{\mu}\m)$.}
\label{fig-1}
\vspace{-2mm}
\end{figure}

In Fig. \ref{fig-1}, we plot the expected number of molecules observed/absorbed by the RX versus time due to the continuous emission of one TX. The analytical curves of Case i), Cases ii)-iii), and Cases iv)-vi) are obtained by \eqref{point1}, \eqref{point-circle,self}, and \eqref{point1,ab}, respectively. We see that the channel responses reach the steady state when the time goes to infinity. We also see the channel response decreases as the TX-RX distances increase.

\section{Proof of Theorem \ref{Theorem:continuous,any}, Theorem \ref{Theorem:continuous,any}, and Theorem \ref{Theorem:continuous,any,ab}}\label{Proof of CR}

The asymptotic channel response due to continuous emission $\overline{N}\left(\vec{x}_i|\vec{x}_j\right)$ in \eqref{point1} can be obtained by multiplying the impulse channel response for a passive RX \cite[Eq.~(8)]{6949026} $\frac{\frac{4}{3}\pi R_0^3}{(4\pi D_{\QS} t)^{3/2}}\exp\left(-\frac{|\vec{x}_i -\vec{x}_j|^{2}}{4D_{\QS} t}-k_{\QS}t\right)$, by the emission mean rate $q_{\QS}$ and then integrating over all time to infinity.

{To obtain (2), we first resort to \cite[Eq. (27)]{Noel2013} to evaluate the impulse channel response of QS molecules observed at the passive RX at time $t$ as
\begin{align}\label{general prob}
\overline{N}\left(\vec{x}_i,t|\vec{x}_j\right)=&\;\Bigg(\frac{1}{2}\left[\erf\left(\tau_{1}\right)+\erf\left(\tau_{2}\right)\right]
-\frac{\sqrt{D_{\QS}t}}{|\vec{x}_i -\vec{x}_j|\sqrt{\pi}}\nonumber\\
&\;\left[\exp\left(-\tau_{1}^{2}\right)-\exp\left(-\tau_{2}^{2}\right)\right]\Bigg)\exp(-k_{\QS}t),
\end{align}
where $\tau_{1}=\frac{R_0+|\vec{x}_i -\vec{x}_j|}{2\sqrt{D_{\QS}t}}$,  $\tau_{2}=\frac{R_0-|\vec{x}_i -\vec{x}_j|}{2\sqrt{D_{\QS}t}}$. We note that \eqref{general prob} is accurate for any value of $|\vec{x}_i -\vec{x}_j|$. We then multiply \eqref{general prob} by the emission rate $q_{\QS}$ and then integrating over all time to infinity to obtain the asymptotic channel response due to continuous emission as
\begin{align}\label{general prob,asy}
&\;\overline{N}\left(\vec{x}_i|\vec{x}_j\right)\nonumber\\
=&\;\frac{q_{\QS}}{2k_{\QS}^2}\left(2k_{\QS}-\exp\left(-\sqrt{\frac{k_{\QS}}{D_{\QS}}}(R_0+|\vec{x}_i -\vec{x}_j|)\right)\Xi\right),
\end{align}
where
\begin{align}\label{xi}
\Xi = \left(-1+\exp\left(2 |\vec{x}_i -\vec{x}_j|\sqrt{\frac{k_{\QS}}{D_{\QS}}}\right)\right)\frac{\left(\sqrt{D_{\QS}k_{\QS}}+k_{\QS}R_0\right)}{|\vec{x}_i -\vec{x}_j|}.
\end{align} 
We note that $\Xi$ goes to $\frac{0}{0}$ in the limit of $|\vec{x}_i -\vec{x}_j|\rightarrow0$. The channel response $\overline{N}_{\self}$ in (2) can be obtained by applying L'H\^{o}pital's rule in the limit of $|\vec{x}_i -\vec{x}_j|\rightarrow0$ to $\Xi$ in \eqref{general prob,asy}.}

The channel response $\overline{h}\left(\vec{x}_\ab|\vec{x}_i\right)$ in \eqref{point1,ab} can be obtained by multiplying the hitting rate of molecules at an absorbing RX \cite[Eq.~(22)]{6807659} $\frac{R_\ab}{|\vec{x}_\ab-\vec{x}_i|}\frac{1}{\sqrt{4\pi D_{\D} t}}\frac{|\vec{x}_\ab-\vec{x}_i|-R_\ab}{t}\exp\left(-\frac{\left(|\vec{x}_\ab-\vec{x}_i|-R_\ab\right)^2}{4 D_{\D}t}\right)$, by $\exp\left(-k_{\QS}t\right)$ and the emission rate $q_{\QS}$, and then integrating over all time to infinity.
\vspace{-2mm}

\section{Proof of Lemma \ref{lem:prob}}\label{proof:Laplace}
We note that ${N}_{\agg}^{\dag}(\vec{x}_i|\lambda)$ is the instantaneous observation at the $i$th B-NM and $\overline{N}_{\agg}^{\dag}(\vec{x}_i|\lambda)$ is its expected observation for a given instantaneous realization of random B-NMs locations. For tractability, we assume the instantaneous number of received molecules, ${N}_{\agg}^{\dag}(\vec{x}_i|\lambda)$, is a Poisson RV with mean $\overline{N}_{\agg}^{\dag}(\vec{x}_i|\lambda)$. By doing so, we have
\begin{align}\label{prob,Laplace1}
&\;{\E}_{\Phi}\Big\{\prob\left(N_{\agg}^{\dag}(\vec{x}_i|\lambda)\geq\eta|\phi(\lambda)\right)\Big\}\nonumber\\
= &\;1-{\E}_{\Phi}\Bigg\{\sum_{n=0}^{\eta-1}\frac{1}{n!}\exp\left\{-\overline{N}_{\agg}^{\dag}(\vec{x}_i|\lambda)\right\}\left(\overline{N}_{\agg}^{\dag}(\vec{x}_i|\lambda)\right)^{n}\Bigg\}\nonumber\\
= &\;1-\sum_{n=0}^{\eta-1}\frac{1}{n!}{\E}_{\Phi}\Bigg\{\exp\left\{-\overline{N}_{\agg}^{\dag}(\vec{x}_i|\lambda)\right\}\left(\overline{N}_{\agg}^{\dag}(\vec{x}_i|\lambda)\right)^{n}\Bigg\}.
\end{align}

We apply $\exp\left\{-\overline{N}_{\agg}^{\dag}(\vec{x}_i|\lambda)\right\}\left(\overline{N}_{\agg}^{\dag}(\vec{x}_i|\lambda)\right)^{n}=\frac{\partial^n\exp\left\{\overline{N}_{\agg}^{\dag}(\vec{x}_i|\lambda)\rho\right\}}{\partial \rho^n}\Bigg|_{\rho=-1}$ in \cite{8030318} to derive \eqref{prob,Laplace1} as
\begin{align}\label{prob,Laplace2}
&\;{\E}_{\Phi}\Big\{\prob\left(N_{\agg}^{\dag}(\vec{x}_i|\lambda)\geq\eta|\phi(\lambda)\right)\Big\}\nonumber\\
= &\;1-\sum_{n=0}^{\eta-1}\frac{1}{n!}{\E}_{\Phi}\Bigg\{\frac{\partial^n\exp\left\{\overline{N}_{\agg}^{\dag}(\vec{x}_i|\lambda)\rho\right\}}{\partial \rho^n}\Bigg|_{\rho=-1}\Bigg\}\nonumber\\
= &\;1-\sum_{n=0}^{\eta-1}\frac{1}{n!}\int_{\tau=0}^{\infty}\frac{\partial^n\exp\{\tau \rho\}}{\partial \rho^n}\Bigg|_{\rho=-1}f(\overline{N}_{\agg}^{\dag}(\vec{x}_i|\lambda)=\tau)\,d\tau,
\end{align}
where $f(\overline{N}_{\agg}^{\dag}(\vec{x}_i|\lambda)=\tau)$ denotes the PMF of $\overline{N}_{\agg}^{\dag}(\vec{x}_i|\lambda)$. By exchanging the order of derivative and integral, we rewrite \eqref{prob,Laplace2} as
\begin{align}\label{prob,Laplace3}
&\;{\E}_{\Phi}\Big\{\prob\left(N_{\agg}^{\dag}(\vec{x}_i|\lambda)\geq\eta|\phi(\lambda)\right)\Big\}\nonumber\\
=  &\;1-\sum_{n=0}^{\eta-1}\frac{1}{n!}\frac{\partial^n \int_{\tau=0}^{\infty}\exp\{\tau \rho\}f(\overline{N}_{\agg}^{\dag}(\vec{x}_i|\lambda)=\tau)\,d\tau}{\partial \rho^n}\Bigg|_{\rho=-1}\nonumber\\
= &\;1-\sum_{n=0}^{\eta-1}\frac{1}{n!}\frac{\partial^n {\E}_{\Phi}\Big\{\exp\left\{\overline{N}_{\agg}^{\dag}(\vec{x}_i|\lambda)\rho\right\}\Big\}}{\partial \rho^n}\Bigg|_{\rho=-1}\nonumber\\
= &\;1-\sum_{n=0}^{\eta-1}\frac{1}{n!}\frac{\partial^n \mathcal{L}_{\overline{N}_{\agg}^{\dag}(\vec{x}_i|\lambda)}(-\rho)}{\partial \rho^n}\Bigg|_{\rho=-1},
\end{align}
where $\mathcal{L}_{\overline{N}_{\agg}^{\dag}(\vec{x}_i|\lambda)}(\cdot)$ is the Laplace transform of $\overline{N}_{\agg}^{\dag}(\vec{x}_i|\lambda)$, which is defined as $\mathcal{L}_{\overline{N}_{\agg}^{\dag}(\vec{x}_i|\lambda)}(s)
= {\E}_{\Phi}\Big\{\exp\left\{-s\overline{N}_{\agg}^{\dag}(\vec{x}_i|\lambda)\right\}\Big\}$.
We next derive $\mathcal{L}_{\overline{N}_{\agg}^{\dag}(\vec{x}_i|\lambda)}(s)$. We first recall that the $i$th B-NM observes molecules in the environment released from all B-NMs (also including the molecules released from itself). Thus, we have
\begin{align}\label{obsExp}
&\;\overline{N}_{\agg}^{\dag}\left(\vec{x}_i|\lambda\right)=\sum_{\vec{x}_j\in\Phi\left(\lambda\right)}\overline{N}\left(\vec{x}_i|\vec{x}_j\right)
\nonumber\\
= &\;\overline{N}\left(\vec{x}_i|\vec{x}_i\right)+\sum_{\vec{x}_j\in\Phi\left(\lambda\right)/\vec{x}_i}\overline{N}\left(\vec{x}_i|\vec{x}_j\right)
\nonumber\\
= &\;\overline{N}_{\self}+\sum_{\vec{a}\in\Phi\left(\acute{\lambda}\right)}\overline{N}\left(\vec{x}_i|\vec{a}\right),
\end{align}
where $\overline{N}_{\self}$ is given in \eqref{point-circle,self}
and $\acute{\lambda}={\left(\lambda\frac{4}{3}\pi R_1^3-1\right)}/{\frac{4}{3}\pi R_1^3}$. We consider a new density $\acute{\lambda}$ to keep the average number of bacteria the same after the approximation of \eqref{obsExp}. We then apply \eqref{obsExp} to $\mathcal{L}_{\overline{N}_{\agg}^{\dag}(\vec{x}_i|\lambda)}(s)$ to rewrite it as
\begin{align}\label{Laplace}
&\;\mathcal{L}_{\overline{N}_{\agg}^{\dag}(\vec{x}_i|\lambda)}(s)\nonumber\\
=&\;{\E}_{\Phi}\Bigg\{\exp\Bigg\{-s \Bigg\{\sum_{\vec{a}\in\Phi(\acute{\lambda})}\overline{N}(\vec{x}_i|\vec{a})+\overline{N}_{\self}\Bigg\} \Bigg\}\Bigg\},\nonumber\\
= &\;{\E}_{\Phi}\Bigg\{\exp\Bigg\{-s \Bigg\{\sum_{\vec{a}\in\Phi(\acute{\lambda})}\overline{N}(\vec{x}_i|\vec{a})\Bigg\} \Bigg\}\exp\left\{-s \overline{N}_{\self}\right\}\Bigg\}\nonumber\\
= &\;\exp\left(-s\overline{N}_{\self}\right){\E}_{\Phi}\Bigg\{\prod_{\vec{a}\in\Phi(\acute{\lambda})}\exp\left\{-s \overline{N}(\vec{x}_i|\vec{a})\right\}\Bigg\}.
\end{align}

Using the probability generating functional (PGFL) for the PPP \cite[eq.~(4.8)]{Haenggi:2012:SGW:2480878}, we rewrite \eqref{Laplace} as
\begin{align}\label{Laplace11}
&\;\mathcal{L}_{\overline{N}_{\agg}^{\dag}(\vec{x}_i|\lambda)}(s)
=\exp\Bigg\{-s\overline{N}_{\self}-\acute{\lambda}\int_{|\vec{r}|=0}^{R_1}\int_{\theta=0}^{2\pi}\int_{\varphi=0}^{\pi}\nonumber\\
&\;\times\left(1-\exp\left(-s\overline{N}(\vec{x}_i|\vec{r})\right)\right)|\vec{r}|^2 \sin\varphi\,d|\vec{r}|\,d\theta\,d\varphi \Bigg\}.
\end{align}

Applying \eqref{point1} to \eqref{Laplace11}, we rewrite \eqref{Laplace11} as 
\begin{align}\label{Laplace2}
&\;\mathcal{L}_{\overline{N}_{\agg}^{\dag}(\vec{x}_i|\lambda)}(s)
=\exp\Bigg\{-s\overline{N}_{\self}-\acute{\lambda}\int_{|\vec{r}|=0}^{R_1}\int_{\theta=0}^{2\pi}\int_{\varphi=0}^{\pi}\nonumber\\
&\;\times\left(1-\exp\left(-s\frac{\exp\left(-\sqrt{\frac{k_{\QS}}{D_{\QS}}}|\vec{x}_i -\vec{r}|\right)q_{\QS} R_0^3}{3D_{\QS}|\vec{x}_i -\vec{r}|}\right)\right)\nonumber\\
&\;\times |\vec{r}|^2 \sin\varphi\,d|\vec{r}|\,d\theta\,d\varphi \Bigg\},
\end{align}
where $|\vec{x}_i -\vec{r}|=\Omega\left(|\vec{x}_i|,|\vec{r}|,\varphi,\theta\right)$. Combining \eqref{prob,Laplace3} and \eqref{Laplace2}, we arrive at \eqref{activation, prob}. This completes the proof.

\section{Proof of Theorem \ref{the:AggDrug}}\label{proof:AggDrug}
To prove \eqref{agg,abRate,app}, we have
\begin{align}\label{agg,abRate,app1}
&\;{\E}_{\Phi}\Big\{\overline{h}_{\agg}\left(\vec{x}_\ab\right)\Big\}\nonumber\\
= &\;{\E}_{\Phi}\Big\{\sum_{\vec{x}_i\in \Phi}\overline{h}\left(\vec{x}_\ab|\vec{x}_i\right)\prob\left(N_{\agg}^{\dag}(\vec{x}_i|\lambda)\geq\eta|\phi(\lambda)\right)\Big\}\nonumber\\
\overset{(a)}{\approx} &\;{\E}_{\Phi}\Big\{\sum_{\vec{x}_i\in \Phi}\overline{h}\left(\vec{x}_\ab|\vec{x}_i\right){\E}_{\Phi}\Big\{\prob\left(N_{\agg}^{\dag}(\vec{x}_i|\lambda)\geq\eta|\phi(\lambda)\right)\Big\}\Big\}\nonumber\\
\overset{(b)}{\approx} &\; \lambda\int_{V_{R_1}}\overline{h}\left(\vec{x}_\ab|\vec{x}\right){\E}_{\Phi}\Big\{\prob\left(N_{\agg}^{\dag}(\vec{x}|\lambda)\geq\eta|\phi(\lambda)\right)\Big\}\,d\vec{x}\nonumber\\
\overset{(c)}{\approx} &\; q_{\D}\lambda\int_{V_{R_1}}
\exp\left(\sqrt{\frac{k_{\D}}{D_{\D}}}(R_\ab-|\vec{x}_\ab-\vec{x}|)\right)\frac{R_\ab}{|\vec{x}_\ab-\vec{x}|}\nonumber\\
&\;\times\left(1-\sum_{n=0}^{\eta-1}\frac{1}{n!}\frac{\partial^n \mathcal{L}_{\overline{N}_{\agg}^{\dag}(\vec{x}|\lambda)}(-\rho)}{\partial \rho^n}\Bigg|_{\rho=-1}\right)\,d\vec{x},
\end{align}
where $V_{R_1}$ is the volume where B-NMs are randomly located. Equality (a) is obtained by applying the approximation $\prob\left(N_{\agg}^{\dag}(\vec{x}_i|\lambda)\geq\eta|\phi(\lambda)\right)\approx{\E}_{\Phi}\Big\{\prob\left(N_{\agg}^{\dag}(\vec{x}_i|\lambda)\geq\eta|\phi(\lambda)\right)\Big\}$. We note that in \eqref{agg,abRate,app1}, $\prob\left(N_{\agg}^{\dag}(\vec{x}_i|\lambda)\geq\eta|\phi(\lambda)\right)$ is the conditional activation probability at
the $i$th B-NM centered at $\vec{x}_i$, for a given instantaneous realization $\phi(\lambda)$. Such an instantaneous activation probability, $\prob\left(N_{\agg}^{\dag}(\vec{x}_i|\lambda)\geq\eta|\phi(\lambda)\right)$, makes solving the exact expression of ${\E}_{\Phi}\Big\{h_{\agg}\left(\vec{x}_\ab\right)\Big\}$ in \eqref{agg,abRate,app1} mathematically intractable, since $\prob\left(N_{\agg}^{\dag}(\vec{x}_i|\lambda)\geq\eta|\phi(\lambda)\right)$ not only depends on $\vec{x}_i$ but also depends on the locations of the other B-NMs in a particular realization of $\phi(\lambda)$. To tackle this problem, we use the expected activation probability over the spatial point process $\Phi$ to approximate the conditional one for a given instantaneous realization $\phi(\lambda)$, i.e., $\prob\left(N_{\agg}^{\dag}(\vec{x}_i|\lambda)\geq\eta|\phi(\lambda)\right)\approx{\E}_{\Phi}\Big\{\prob\left(N_{\agg}^{\dag}(\vec{x}_i|\lambda)\geq\eta|\phi(\lambda)\right)\Big\}$, to solve \eqref{agg,abRate,app1}. This approximation is more accurate when $\lambda$ is lower. This is because when $\lambda$ is lower,
$N_{\agg}^{\dag}(\vec{x}_i|\lambda)$ is closer to ${\E}_{\Phi}\Big\{N_{\agg}^{\dag}(\vec{x}_i|\lambda)\Big\}$. Equality (b) is obtained by applying Campbell's Theorem. Equality (c) is obtained by applying \eqref{point1,ab} and \eqref{activation, prob}. By integrating $\vec{x}$ over $V_{R_1}$ and applying $|\vec{x}_\ab-\vec{x}|=\Lambda(\vec{x},\vec{x}_\ab,\alpha,\beta)$, we obtain \eqref{agg,abRate,app}. This completes the proof.

\section{{Illustration of PPP-distributed B-NMs}}\label{proof:realization}
\begin{figure}[!h]
\centering
\subfigure[Realization 1]{\label{fig5:1}\includegraphics[height=2.2in]{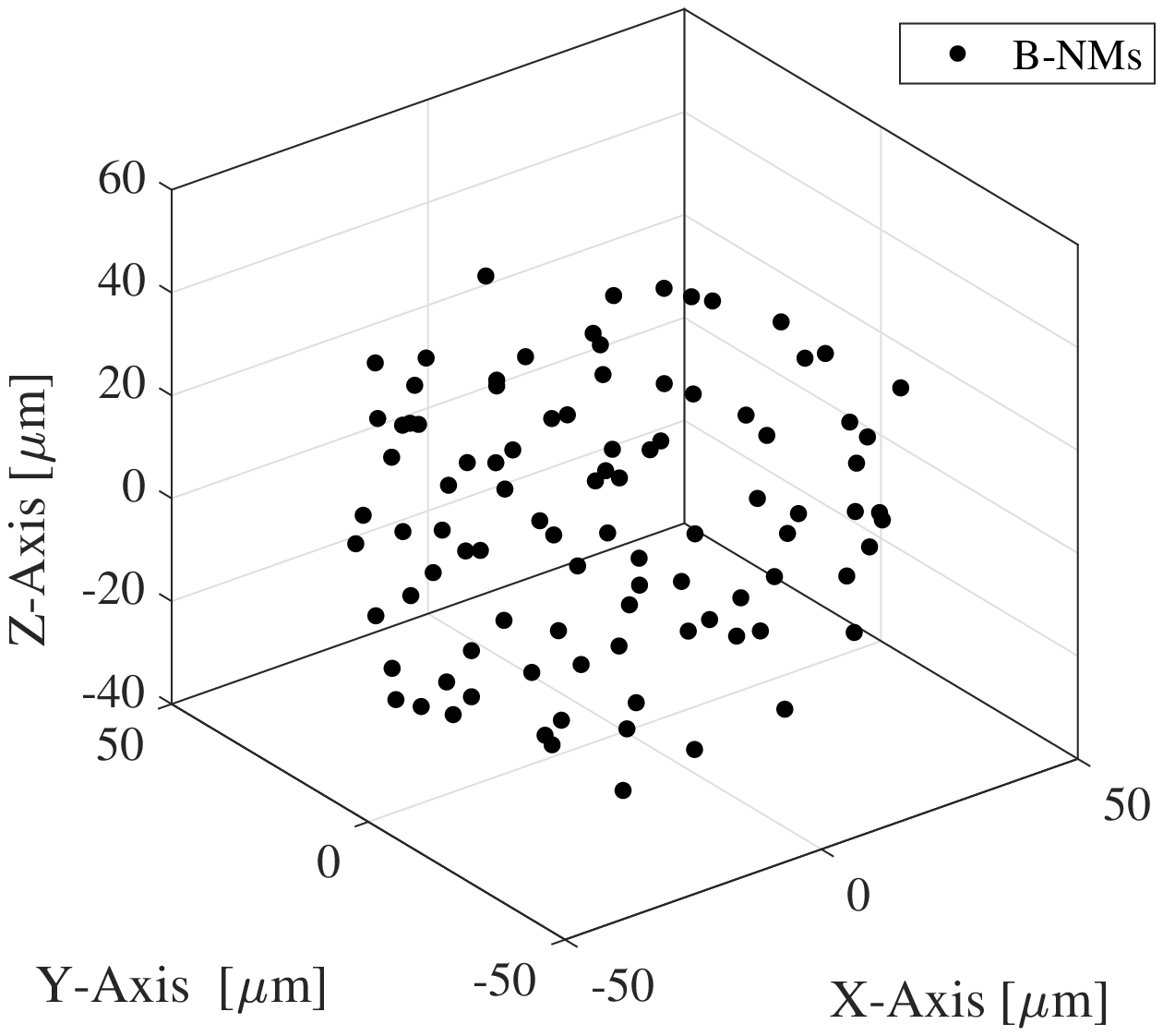}}
\subfigure[Realization 2]{\label{fig5:2}\includegraphics[height=2.2in]{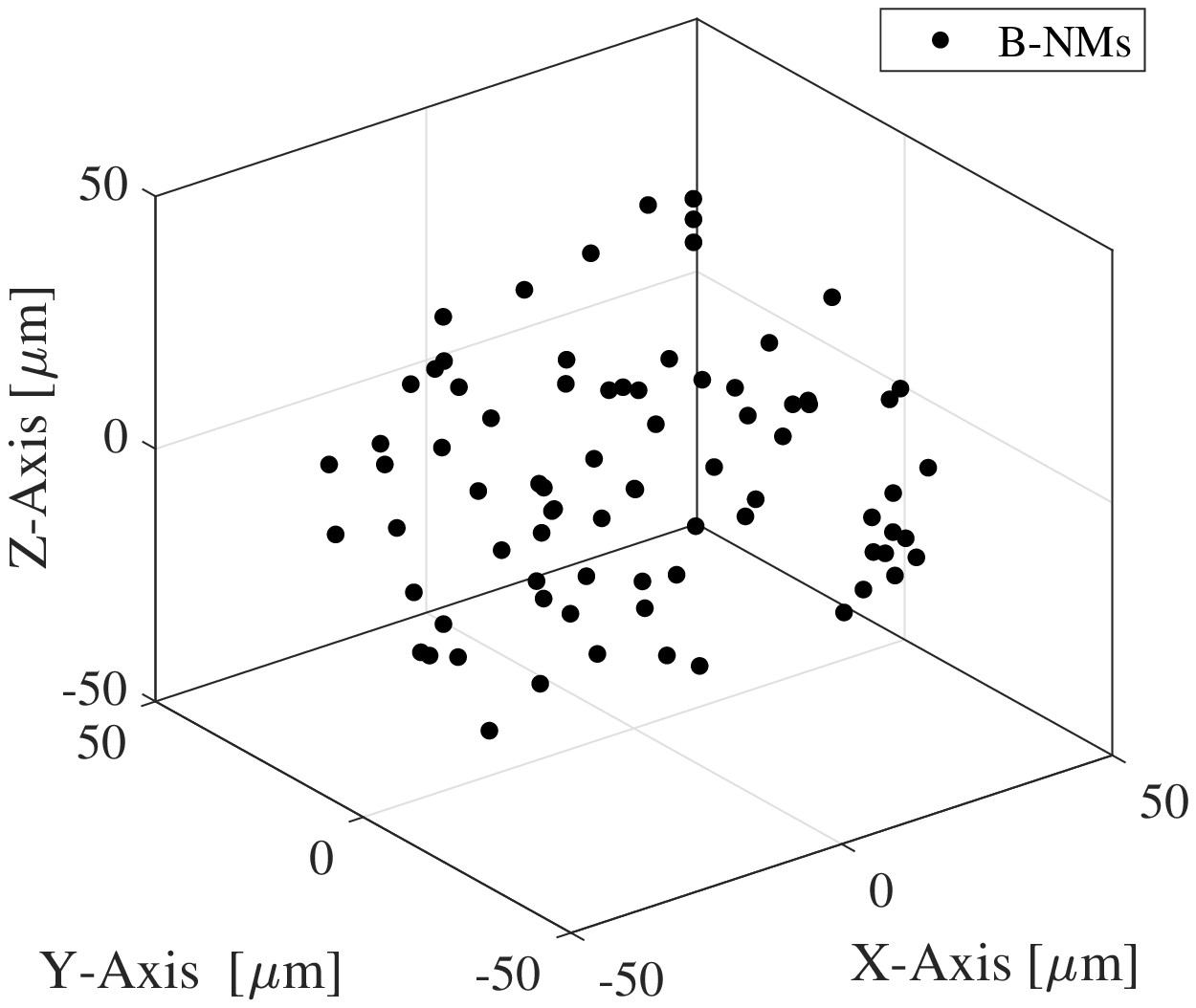}}
\subfigure[Realization 3]{\label{fig5:3}\includegraphics[height=2.2in]{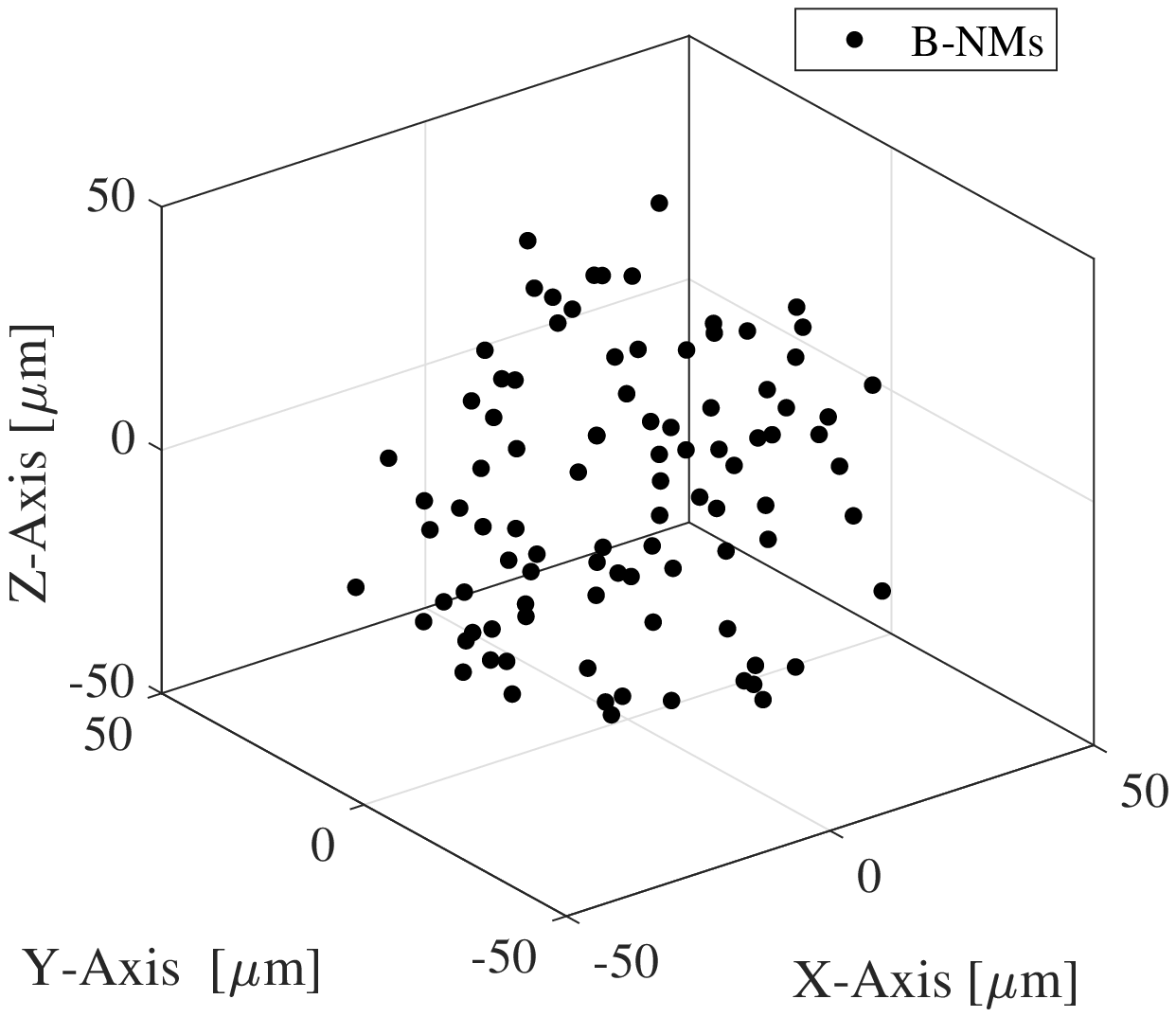}}
\caption{{Three simulation realizations of a 3D PPP of randomly-distributed B-NMs. $R_1=50\,{\mu}\m$, $\lambda=2\times10^{-4}$ B-NMs per ${{\mu}\textrm{m}^{3}}$.}}\label{fig-Re2}
\end{figure}

{Different from the model with deterministic topology, point process (such as PPP considered in our work) models deal with topological randomness. A point process model is abstracted to be a collection of nodes residing in a certain place. The locations of these nodes are not deterministic but subject to uncertainty. In this work, the location of B-NMs are distributed according to a 3D PPP. Three simulation realizations of a 3D PPP of randomly-distributed B-NMs are illustrated in Fig. \ref{fig-Re2}, where the realization is a deterministic set of points \cite{Haenggi:2012:SGW:2480878}.}

\end{appendices}

\end{document}